\documentclass[11pt,prd,preprintnumbers,nofootinbib,superscriptaddress,notitlepage]{revtex4-1}

\usepackage{graphics}
\usepackage{cancel}
\usepackage{color}
\usepackage{xcolor}
\usepackage{multirow}
\usepackage{slashed}
\usepackage{array}
\usepackage{amssymb}
\usepackage{bbold}
\usepackage{graphicx}
\usepackage{mathrsfs}
\usepackage{diagbox}
\usepackage{amsmath}
\usepackage{float}
\usepackage{extarrows}
\usepackage{booktabs}
\usepackage[colorlinks,citecolor=blue,anchorcolor=red,menucolor=red,linkcolor=red,filecolor=red,runcolor=red,urlcolor=red,frenchlinks=red]{hyperref}
\usepackage{subfigure}
\usepackage{ulem}
\usepackage{braket}

\usepackage{bm}

\makeatletter
\newcommand{\thickhline}{\noalign {\ifnum 0=`}\fi \hrule height 1pt\futurelet \reserved@a \@xhline}
\newcolumntype{"}{@{\hskip\tabcolsep\vrule width 1pt\hskip\tabcolsep}}
\makeatother

\begin{document}
\title{Spin correlations and quantum entanglement in $\gamma\gamma \to t\bar t$ at polarized photon colliders with NLO QCD corrections}

\author{Jun Jiang}
\email{jiangjun87@sdu.edu.cn}
\affiliation{School of Physics, Shandong University, Jinan, Shandong 250100, China}

\author{Zong-Guo Si}
\email{zgsi@sdu.edu.cn}
\affiliation{School of Physics, Shandong University, Jinan, Shandong 250100, China}

\author{Han Zhang}
\email{han.zhang@mail.sdu.edu.cn}
\affiliation{School of Physics, Shandong University, Jinan, Shandong 250100, China}

\author{Xin-Yi Zhang}
\email{xinyizhang@mail.sdu.edu.cn}
\affiliation{School of Physics and Technology, University of Jinan, Jinan, Shandong 250022, China}

\begin{abstract}
We study spin correlations and quantum entanglement in the process $\gamma\gamma\to t\bar t$  with the photons coming from Compton backscattered laser beam.  
We present predictions for the  cross sections and spin observables including next-to-leading order (NLO) QCD corrections under various beam-polarization configurations. 
The NLO QCD corrections significantly enhance the total cross section while having only a minor impact on spin observables. Using the spin density matrix of the $t\bar t$ system, we further investigate quantum entanglement and Bell nonlocality. 
We find that the entanglement is the strongest near the $t\bar t$ invariant-mass threshold, while both entanglement and Bell nonlocality are highly sensitive to the initial beam polarization. 
Our results provide a theoretical basis for future studies of quantum correlations in top-quark pair production at photon colliders.

\end{abstract}
\maketitle

\section{Introduction}
\label{sec:introduction}

As the heaviest elementary fermion in the Standard Model (SM), the top quark plays a special role in  studies of the electroweak sector, and its large Yukawa coupling suggests a close connection to the electroweak symmetry breaking mechanism~\cite{Bernreuther:2008ju, Beneke:2000hk, Chakraborty:2003iw}.
Moreover, the top quark decays before hadronization takes place, so that its spin information is preserved in the angular distributions of its decay products~\cite{Bigi:1986jk}. 
The polarization and spin correlations of top-quark pairs therefore provide sensitive probes of the production dynamics~\cite{Mahlon:2010gw, Mahlon:1997uc}, higher-order corrections~\cite{Bernreuther:2004jv, Bernreuther:2006vg, Bernreuther:2001rq}, and possible new physics contributions~\cite{Bernreuther:2013aga, Atwood:1992vj, Aguilar-Saavedra:2008nuh}. 
Experimentally, spin correlation observables have been measured at the LHC through angular distributions of top decay products~\cite{ATLAS:2014abv, CMS:2019nrx}.

The bipartite spin system formed by the spin degree of freedom of top quark pair allows us to investigate the direct implications of quantum theory such as entanglement~\cite{Gigena:2017agv}. 
Entanglement properties of the $t\bar t$ system can be characterized through observables, such as concurrence and Bell parameters~\cite{Fabbrichesi:2021npl, Fabbrichesi:2025psr, Antozzi:2026vdi}, and also provide sensitive probes of the production mechanism and the spin configuration of top-quark pairs at the LHC~\cite{Afik:2020onf,ATLAS:2023fsd,CMS:2024pts}.

Compared with hadron colliders, photon colliders provide a clean environment for studying top-quark spin correlations and entanglement observables. 
At a high-energy linear lepton collider, photon beams can be obtained via Compton backscattering of laser photons off electron beams, where the photon energy spectrum and polarization can be tuned through the polarizations of the electron and laser beams~\cite{Ginzburg:1981vm, LinearCollider:2025lya, LinearColliderVision:2025hlt, Ginzburg:1982yr}. 
Since photons couple directly to the top quark, the process $\gamma\gamma \to t\bar t$ is highly sensitive to the initial photon helicities. 
In particular, near the production threshold, the initial-state polarization can strongly affect the spin configuration of the $t\bar t$ pair, making this process especially suitable for studies of quantum entanglement~\cite{HARLANDER1995137, Parke:1996pr}. 
Furthermore, photon colliders are complementary to $e^+e^-$ colliders in terms of production mechanisms and helicity structures, providing additional information on top-quark pair production dynamics~\cite{Bigi:1992pr}.

For precision studies at colliders, leading-order (LO) predictions for spin correlations and entanglement observables of top-quark pair have been investigated~\cite{Mahlon:1995zn, Afik:2022dgh, Choi:2026omc, Barr:2024djo, Afik:2022kwm}.
Higher-order corrections to $\gamma\gamma \to t\bar t$, including NLO QCD and electroweak corrections to the total cross section, as well as threshold-enhanced effects and  next-next-to-leading order (NNLO) QCD contributions, have been investigated~\cite{Jikia:1996bi,Melles:1998xd, Denner:1995ar,Penin:1998mx,Capatti:2025khs}.
However, the impact of  NLO QCD corrections on spin-dependent observables, particularly quantum entanglement measures, remains  unexplored, especially in the presence of polarized photon beams.
Therefore, a systematic study of the NLO QCD corrections to spin correlation and entanglement observables in $\gamma\gamma \to t\bar t$ production  is necessary.
In this work, we study the spin correlations and entanglement observables in $\gamma\gamma \to t\bar t$ production at a photon collider, including  NLO QCD corrections. 
High-energy photon beams are obtained via Compton backscattering of laser photons off electron beams, and the polarization of the initial photons is  taken into account. 

The structure of this paper is as follows. 
In Section \ref{sec2}, we introduce the theoretical framework for $\gamma\gamma \to t \bar{t}$ through Compton backscattering, reviewing the spin density matrix of $t \bar{t}$ production and the associated spin correlations.
In Section \ref{sec4}, we present the numerical results for the cross section as well as spin correlation effects corresponding to different polarization configurations of initial photons. 
We also discuss the prospects of probing quantum entanglement through this process.
Section \ref{sec5} is reserved for the summary.


\section{Theoretical Framework}
\label{sec2}
In a photon collider setup, high-energy photon beams are generated through Compton backscattering of laser photons off energetic electron beams~\cite{  Ginzburg:1982yr}. 
Consequently, the observable cross section for top-quark pair production is obtained by convoluting the partonic cross section with the energy spectra of the backscattered photons,
\begin{equation}
\sigma_{t\bar t}=
\int_0^{y_{\rm max}}dy_1
\int_0^{y_{\rm max}}dy_2\,
f_\gamma^e(y_1,P_e^{(1)},P_L^{(1)})
f_\gamma^e(y_2,P_e^{(2)},P_L^{(2)})
\,
\hat{\sigma}_{\gamma\gamma\to t\bar t X}
(\hat s,\lambda_1,\lambda_2),
\end{equation}
where $y_i$ denotes the fraction of the electron-beam energy carried by the $i$th backscattered photon, and $\hat s=y_1y_2s_{ee}$ is the invariant mass squared of the photon--photon subsystem.

The normalized energy spectrum of the backscattered photons is determined by the Compton scattering dynamics and is given by~\cite{Ginzburg:1981vm}
\begin{equation}
f_\gamma^e(y,P_e,P_L)
=
{\cal N}^{-1}
\left[
\frac{1}{1-y}
-y
+(2\xi-1)^2
-
P_eP_L\,x\,\xi(2\xi-1)(2-y)
\right],
\end{equation}
with
\begin{equation}
\xi=\frac{y}{x(1-y)},
\qquad
x=\frac{4E_LE_e}{m_e^2}.
\end{equation}
Here, $E_e$ and $E_L$ denote the energies of the electron beam and laser beam, respectively, while $m_e$ is the electron mass. 
The quantities $P_e$ and $P_L$ represent the polarization of the electron beam and laser photons, respectively. 
We consider $|P_e|=0.85$ and $|P_L|=1$ for the polarized configurations, consistent with the beam polarization parameters considered for photon colliders~\cite{Brau:2007zza}.
The normalization factor ${\cal N}$ satisfies
\begin{equation}
\int_0^{y_{\rm max}}
f_\gamma^e(y,P_e,P_L)\,dy=1.
\end{equation}
The kinematically allowed range of the photon energy fraction is
$0\le y\le x/(x+1)$.
To avoid $e^+e^-$ pair creation in the interaction between the backscattered photons and laser photons, the parameter $x$ is usually limited to $x_{\rm max}=2(1+\sqrt2)$. 
This corresponds to an optimal laser energy of $E_L\simeq1.26~{\rm eV}$ for $E_e=250~{\rm GeV}$.

The Compton backscattering dynamics also determines the polarization of the scattered photons.
The  polarization of the incoming photons entering the hard-scattering subprocess is defined by
\begin{equation}
\lambda_i=
P_\gamma(y_i,P_e^{(i)},P_L^{(i)}),
\qquad i=1,2,
\end{equation}
where the function $P_\gamma(y,P_e,P_L)$ describes the polarization degree of a backscattered photon carrying an energy fraction $y$~\cite{Ginzburg:1982yr}:
\begin{equation}
P_\gamma(y,P_e,P_L)=
\frac{1}
{{\cal N}f_\gamma^e(y,P_e,P_L)}
\left\{
x\,\xi P_e
\left[
1+(1-y)(2\xi-1)^2
\right]
-(2\xi-1)P_L
\left[
\frac{1}{1-y}+1-y
\right]
\right\}.
\end{equation}

For the hard-scattering subprocess, we consider
\begin{equation}\label{eq:7}
\gamma(p_1,\lambda_1)+\gamma(p_2,\lambda_2)
\to t(k_1,s_t)+\bar{t}(k_2,s_{\bar t}),
\end{equation}
including NLO QCD corrections and spin correlations,
where $s_t$ and $s_{\bar{t}}$ denote the spin four-vectors of the top quark and antiquark respectively, and  satisfy the following relation:
\begin{equation}
s_t^2=s_{\bar{t}}^2=-1, \qquad
k_1\cdot s_t = k_2\cdot s_{\bar{t}}=0.
\end{equation}

The polarized partonic differential cross section is given by
\begin{equation}
d\hat{\sigma}(\lambda_1,\lambda_2,s_t,s_{\bar{t}})
= \frac{N_c}{2\hat{s}} |M_0|^2 d\Phi,
\end{equation}
where $N_c$ is the number of colors, $d\Phi$ is two-body phase space.
The spin dependence of the squared matrix element $ |M_0|^2$ can be decomposed into
unpolarized, single-spin polarization, and spin correlation contributions,
which are conveniently described within the spin density matrix formalism
introduced below.
At NLO, both virtual and real gluon contributions are included.
These corrections affect not only the production rate but also the spin structure of the $t\bar t$ system.

Let's denote the amplitude for the process in Eq.~(\ref{eq:7}) by
\begin{equation}
    \mathcal{M}^{\alpha,\beta,\lambda_1,\lambda_2} =
    \bra{t\left(k_1, \alpha\right),\bar{t}\left(k_2, \beta\right)}
        \mathcal{M}\ket{\gamma\left(p_1, \lambda_1),\gamma\left(p_2, \lambda_2\right)\right)},
\end{equation}
where $\alpha$, $\beta$ are spin labels referring to $t$ and $\bar{t}$, respectively. In perturbative theory, $\mathcal{M}^{\alpha,\beta,\lambda_1,\lambda_2}$ permits an expansion in the electroweak coupling $e$ and the strong coupling $g_s$. Let's further denote the $\mathcal{O}\left(e^2\right)$ term in this expansion by
$\mathcal{M}_{LO}^{\alpha,\beta,\lambda_1,\lambda_2}$, and the $\mathcal{O}\left(e^2g_s^2\right)$ term by $\mathcal{M}_{V}^{\alpha,\beta,\lambda_1,\lambda_2}$.
Then, the LO contribution to the production density matrix is given by
\begin{equation}
    R_{LO}^{\alpha_1\alpha_2,\beta_1\beta_2} = \left(\mathcal{M}_{LO}^{\alpha_2,\beta_2,\lambda_1,\lambda_2}\right)^{*}
    \mathcal{M}_{LO}^{\alpha_1,\beta_1,\lambda_1,\lambda_2},
    \label{eq:DenMatrixLO}
\end{equation}
which is at $\mathcal{O}\left(\alpha^2\right)$. Similarly, the contribution
from one loop QCD correction, which is at $\mathcal{O}\left(\alpha^2\alpha_s\right)$,
is given by
\begin{equation}
    R_{V}^{\alpha_1\alpha_2,\beta_1\beta_2} = \left(\mathcal{M}_{V}^{\alpha_2,\beta_2,\lambda_1,\lambda_2}\right)^{*}
    \mathcal{M}_{LO}^{\alpha_1,\beta_1,\lambda_1,\lambda_2} +
    \left(\mathcal{M}_{LO}^{\alpha_2,\beta_2,\lambda_1,\lambda_2}\right)^{*}
    \mathcal{M}_{V}^{\alpha_1,\beta_1,\lambda_1,\lambda_2}.
    \label{eq:DenMatrixV}
\end{equation}
At $\mathcal{O}\left(\alpha^2\alpha_s\right)$, the process with the emission
of an additional gluon
\begin{equation}
\gamma(p_1,\lambda_1)+\gamma(p_2,\lambda_2)
\to t(k_1,s_t)+\bar{t}(k_2,s_{\bar t}) + g\left(k_3, \lambda_3\right)
\label{eq:real-emission-process}
\end{equation}
contributes as well. Here $k_3$ and $\lambda_3$ refer to the momentum and helicity
of the final state gluon, respectively. Let's denote the corresponding $\mathcal{O}\left(e^2g_s\right)$ amplitude by $\mathcal{M}_R^{\alpha,\beta,\lambda_3,\lambda_1,\lambda_2}$. Since we are only interested in the polarization
and correlation between $t$ and $\bar{t}$, the real contribution to the production density matrix should be defined with the helicity of the final state gluon summed over:
\begin{equation}
    R_{R}^{\alpha_1\alpha_2,\beta_1\beta_2} =
    \sum_{\lambda_3}\left(\mathcal{M}_{R}^{\alpha_2,\beta_2,\lambda_3,\lambda_1,\lambda_2}\right)^{*}
    \mathcal{M}_{R}^{\alpha_1,\beta_1,\lambda_3,\lambda_1,\lambda_2}.
    \label{eq:DenMatrixR}
\end{equation}
Thus, after summing over the gluon helicity, the real-emission contribution $R_R$, as well as the Born and virtual contributions $R_{LO}$ and $R_V$,
are all $(2\times2)\otimes(2\times2)$ matrices acting on the $t \bar{t}$ spin space.
Therefore, the complete production spin density matrix can be expanded in the Pauli basis as~\cite{Bernreuther:2015yna}

\begin{equation}
R
=
\tilde A\,\mathbb{1}\otimes\mathbb{1}
+
\tilde B_i^{+}\,\sigma^i\otimes\mathbb{1}
+
\tilde B_i^{-}\,\mathbb{1}\otimes\sigma^i
+
\tilde C_{ij}\,\sigma^i\otimes\sigma^j.
\end{equation}
Here the scalar coefficient $\tilde A$ denotes the unpolarized production rate, while the vectors $\tilde B_i^{\pm}$ describe the single-particle polarization of the top and anti-top quarks, and the tensor $\tilde C_{ij}$ represents the spin correlations between top and anti-top quarks.
The polarization and correlation coefficients are related to the spherical-harmonic expansion coefficients (up to \(l = 1\)) of the angular distributions of the top quark decay products. These coefficients are affected by higher-order QCD corrections and realistic acceptance cuts. At NLO, the gluon radiation introduces additional kinematic degrees of freedom and may require higher order spherical harmonics. The cancellation of infrared singularities ensures the finiteness of the corresponding NLO observables, but the coefficients remain sensitive to acceptance cuts, which are not included in the present analysis.

To analyze the kinematic structure of these coefficients, we work in the $t\bar t$ zero-momentum frame (ZMF).
We denote the direction of flight of top quark in this frame by $\hat{\mathbf k}$ and that of incoming photon by $\hat{\mathbf p}$~\cite{Bernreuther:2015yna}, then the right-handed orthonormal basis $\{\hat{\mathbf r},\hat{\mathbf k},\hat{\mathbf n}\}$ can be defined as: 
\begin{equation}
\hat{\mathbf r}=\frac{1}{r}(\hat{\mathbf p}-z\hat{\mathbf k}),
\qquad
\hat{\mathbf n}=\frac{1}{r}(\hat{\mathbf p}\times\hat{\mathbf k}),
\end{equation}
with
\begin{equation}
z=\hat{\mathbf p}\cdot\hat{\mathbf k},
\qquad
r=\sqrt{1-z^2}.
\end{equation}
In this basis, the polarization vectors and spin correlation tensor can be expanded as
\begin{align}
\tilde B_i^\pm &= b_r^\pm \hat r_i + b_k^\pm \hat k_i + b_n^\pm \hat n_i ,\\
\tilde C_{ij} &=
c_{rr}\hat r_i\hat r_j + c_{kk}\hat k_i\hat k_j + c_{nn}\hat n_i\hat n_j
+ c_{rk}(\hat r_i\hat k_j+\hat k_i\hat r_j)
+ c_{rn}(\hat r_i\hat n_j+\hat n_i\hat r_j)
+ c_{kn}(\hat k_i\hat n_j+\hat n_i\hat k_j).
\end{align}
The coefficients $b_v^\pm$ and $c_{vv'}$ are scalar functions of the partonic center-of-mass energy $\hat s$ and the cosine of the scattering-angle $z$.

With a slight abuse of notation, we denote by $R$ for both integrated and differential production spin density matrix. 
Up to $\mathcal{O}\left(\alpha^2\alpha_s\right)$, the resulting matrix can be written as

\begin{equation}
R = R_{LO} + R_V + R_R.
\end{equation}
The corresponding normalized density matrix is defined as
\begin{equation}
\rho = \frac{R}{\mathrm{Tr}[R]},
\qquad
\mathrm{Tr}[\rho]=1.
\end{equation}
The normalization removes the overall production-rate factor while retaining the relative spin information relevant for the normalized spin observables and quantum-information measures.
The condition $\mathrm{Tr}[\rho]=1$ fixes only the normalization and does not imply a pure state; the purity is instead characterized by $\mathrm{Tr}[\rho^2]$, which is generally smaller than unity for the mixed $t\bar t$ state considered here.
In terms of the normalized polarization vectors and correlation tensor,
the density matrix can be written as
\begin{equation}
\rho =
\frac14
\left[
\mathbb{1}\otimes\mathbb{1}
+
B_i^+\,\sigma^i\otimes\mathbb{1}
+
B_i^-\,\mathbb{1}\otimes\sigma^i
+
C_{ij}\,\sigma^i\otimes\sigma^j
\right].
\end{equation}
The normalized polarization vectors and correlation tensor are defined as
\begin{equation}
B_i^\pm = \frac{\tilde B_i^\pm}{\tilde A}, 
\qquad
C_{ij} = \frac{\tilde C_{ij}}{\tilde A}.
\end{equation}

To obtain specific spin observables, the polarization vectors and spin correlation tensor are projected onto the chosen spin quantization axes.
In the present analysis, we use the orthonormal basis $\{\hat{\mathbf r},\hat{\mathbf k},\hat{\mathbf n}\}$ introduced above.
The single-particle polarization along an arbitrary spin quantization axis $\hat{\mathbf a}$ is defined as
\begin{equation}
B_a^\pm = 2\left\langle \hat{\mathbf a}\cdot \vec s_{t,\bar t} \right\rangle ,
\end{equation}
where $\vec s_{t,\bar t}$ denotes the spin operator of the top (anti-top) quark. 
Similarly, the spin correlation coefficients are defined by
\begin{equation}
C_{ab}
=
4\left\langle (\hat{\mathbf a}\cdot \vec s_t)(\hat{\mathbf b}\cdot \vec s_{\bar t}) \right\rangle ,
\end{equation}
with $\hat{\mathbf a}$ and $\hat{\mathbf b}$ being the chosen spin-projection axes for the top and anti-top quarks, respectively. 
Equivalently, these observables correspond to projections of the polarization vectors and correlation tensor,
\begin{equation}
B_a^\pm = a_i B_i^\pm,
\qquad
C_{ab} = a_i b_j C_{ij}.
\end{equation}

In the numerical analysis, these quantities are evaluated including NLO QCD corrections.
At NLO, the production density matrix receives contributions from virtual and real gluon emission, which modify the spin-dependent coefficients while leaving the spin operator basis unchanged. 
Consequently, the polarization vectors $B_i^\pm$ and correlation tensor $C_{ij}$ receive corresponding corrections.

As a quantitative measure of entanglement, we employ the concurrence~\cite{Wootters:1997id}
\begin{equation}
\mathcal{C}[\rho] =
\max\left(0, \zeta_1-\zeta_2-\zeta_3-\zeta_4\right),
\end{equation}
where $\zeta_1 \ge \zeta_2 \ge \zeta_3 \ge \zeta_4$, and  $\zeta_i$ are the square roots of the eigenvalues of
\begin{equation}
\mathcal{R}  = \rho\,\left(\sigma_2\otimes\sigma_2\right)\,
\rho^*\,\left(\sigma_2\otimes\sigma_2\right)\,,
\end{equation}
with $\rho^*$ denoting complex conjugation and $\sigma_2$ being the second Pauli matrix.

Nonlocal correlations are tested via the Clauser-Horne-Shimony-Holt  (CHSH) inequality~\cite{Clauser:1969ny, Horodecki:1995nsk}. 
Defining the matrix
\begin{equation}
M = C C^T,
\end{equation}
where $C$ denotes the $3\times3$ correlation matrix formed by the coefficients $C_{ij}$, and letting
$m_1 \ge m_2 \ge m_3$
be the eigenvalues of $M$, the Bell inequality is violated when
\begin{equation}
m_{12} = m_1 + m_2 > 1.
\end{equation}

\section{ Numerical results and analysis}\label{sec4}
In this section, we  present numerical results for the NLO QCD corrections to the total cross section and to spin correlation observables in the process $\gamma\gamma \to t \bar{t}$, considering a photon collider setup based on Compton backscattering, and further discuss the quantum entanglement properties of the system.
The input parameters adopted in our calculations are specified as follows~\cite{ParticleDataGroup:2024cfk}:
the top quark mass $m_t=173 \ \mathrm{GeV}$, the electron mass  $m_e=0.511 \ \mathrm{MeV}$, the fine-structure constant $\alpha^{-1} = 128$, and the strong coupling  $\alpha_s(\mu=m_t) = 0.1 $.

\subsection{Cross Section}

In terms of the total cross sections, different initial-state polarization configurations have a significant impact on the \(t\bar t\) production cross section at \(\sqrt{s}=500~\mathrm{GeV}\) as shown in Table~\ref{tab:sigma}. 
At \(\sqrt{s}=500~\mathrm{GeV}\), the cross sections vary substantially among different polarization configurations. 
For the polarization configuration $(P_e^{(1)}, P_e^{(2)} ; P_L^{(1)}, P_L^{(2)}) = (0.85,0.85;-1,-1)$, the LO cross section is $223.99~\mathrm{fb}$, about three times the unpolarized result, whereas the $(0.85,0.85;1,1)$ and $(0.85,-0.85;1,-1)$ configurations are strongly suppressed.
This behavior indicates that  the cross section is highly sensitive to the initial-state helicity configuration. 
Meanwhile, the NLO QCD corrections at $\sqrt{s}=500~\mathrm{GeV}$ are sizable, with $K$-factors in the range $1.4\mbox{--}1.6$.
As a validation of our calculation, we have compared the NLO cross sections with our previous results for the same process reported in Ref.~\cite{Brandenburg:2005uu}, and find good agreement.
This agreement provides an independent check of our implementation of the NLO QCD corrections to the total cross sections.
\begin{table}[ht]
	\centering
	\setlength{\tabcolsep}{7pt}
	\renewcommand{\arraystretch}{1.0}
	\begin{tabular}{cccccc}
		\hline
		\hline
		&
		Unpolarized
		&
		$(0.85,0.85;1,1)$
		&
		$(0.85,0.85;-1,-1)$
		&
		$(0.85,-0.85;1,-1)$
		&
		$(0.85,-0.85;-1,1)$
		\\
		\hline
		$\sigma_{\mathrm{LO}}$ [fb]
		& 71.29
		& 27.65
		& 223.99
		& 20.53
		& 75.64
		\\
		
		$\sigma_{\mathrm{NLO}}$ [fb]
		& 105.25
		& 44.70
		& 318.52
		& 31.63
		& 107.53
		\\
		
		$K=	\sigma_{\mathrm{NLO}}/\sigma_{\mathrm{LO}}$
		& 1.48
		& 1.62
		& 1.42
		& 1.54
		& 1.42
		\\
		\hline
	\end{tabular}
	\caption{LO and NLO cross sections and corresponding $K$-factors for different beam polarization configurations $(P_e^{(1)}, P_e^{(2)} ; P_L^{(1)}, P_L^{(2)}) $ at $\sqrt{s}=500~\mathrm{GeV}$.}
	\label{tab:sigma}
\end{table}

\subsection{Single-spin Polarization and Spin Correlation}
For the top quark single-spin polarization, all components vanish at LO for the unpolarized initial state, while NLO QCD corrections induce only negligible nonzero contributions.
However, for several polarized configurations, sizable effects appear.
In particular, the $(0.85,0.85;-1,-1)$ configuration yields the largest polarization, reaching $|B_k^\pm|\simeq0.699$, whereas the $(0.85,0.85;1,1)$ configuration gives $|B_k^\pm|\simeq0.396$.
For both the $(0.85,-0.85;-1,1)$ and $(0.85,-0.85;1,-1)$ polarization configurations, the dominant polarization direction is along the $r$-axis, with $|B_r^\pm|\simeq0.562$ and $0.107$, respectively.
NLO QCD corrections to the single-spin observables remain at the per-mille level.


Different initial-state polarization configurations significantly affect the spin correlation structure of the $t\bar t$ system, as summarized in Table~\ref{tab:obs_500}. 
Since the off-diagonal spin correlation components, such as $C_{rn}$, $C_{nk}$, and $C_{rk}$, are either vanishing or numerically negligible in all considered configurations, the following discussion focuses on the diagonal components only.
The unpolarized case already exhibits a strong $k$-axis correlation, with $C_{kk}\simeq -0.78$, indicating dominance of the helicity basis. 
The $(0.85,0.85;1,1)$ and $(0.85,0.85;-1,-1)$ configurations both exhibit strong negative spin correlations, with $C_{kk}\simeq -0.86$ and $-0.98$, respectively.
In particular, $C_{kk}$ in the latter case is close to its maximal negative value, while $C_{rr}$ and $C_{nn}$ remain sizable and negative in both configurations.
By contrast, the $(0.85,-0.85;1,-1)$ configuration is similar to the unpolarized case, whereas the $(0.85,-0.85;-1,1)$ configuration strongly suppresses $C_{kk}$ to $\sim -0.1$ and drives $C_{rr}$ positive, indicating a substantial change in the dominant spin structure.
The NLO QCD corrections are moderate and do not alter the overall spin correlation pattern, modifying the observables only at the few-percent level.

\begin{table}[ht]
	\centering
	\setlength{\tabcolsep}{10pt}
	\renewcommand{\arraystretch}{1.2}
	\begin{tabular}{ccc|cccccccc}
		\hline
        \hline
		& \multicolumn{2}{c|}{Unpolarized}
		& \multicolumn{2}{c}{(0.85,0.85;1,1)}
		& \multicolumn{2}{c}{(0.85,0.85;-1,-1)}
		& \multicolumn{2}{c}{(0.85,-0.85;1,-1)}
		& \multicolumn{2}{c}{(0.85,-0.85;-1,1)}
		\\
		& LO & NLO & LO & NLO & LO & NLO & LO & NLO & LO & NLO \\
		\hline
		$ C_{kk} $  & -0.781 & -0.793 & -0.864  & -0.878  & -0.977 & -0.975 & -0.769  & -0.780 & -0.104  & -0.158  \\
		$ C_{rr} $  & -0.478 & -0.492 & -0.628  & -0.653 & -0.657 & -0.649 & -0.487  & -0.502 & 0.252  & 0.202  \\
		$ C_{nn} $  & -0.618 & -0.626 & -0.716  & -0.731 & -0.672 & -0.666 & -0.636  & -0.645 & -0.317  & -0.350  \\
		\hline
	\end{tabular}
	\caption{Spin correlation for unpolarized and various polarized configurations at $\sqrt{s}=500$ GeV.}
	\label{tab:obs_500}
\end{table}

\subsection{Entanglement Observables}

\begin{figure}[htbp]
	\begin{center}
		\subfigure[]{\label{C_unpol}
			\includegraphics[width=0.4\textwidth]{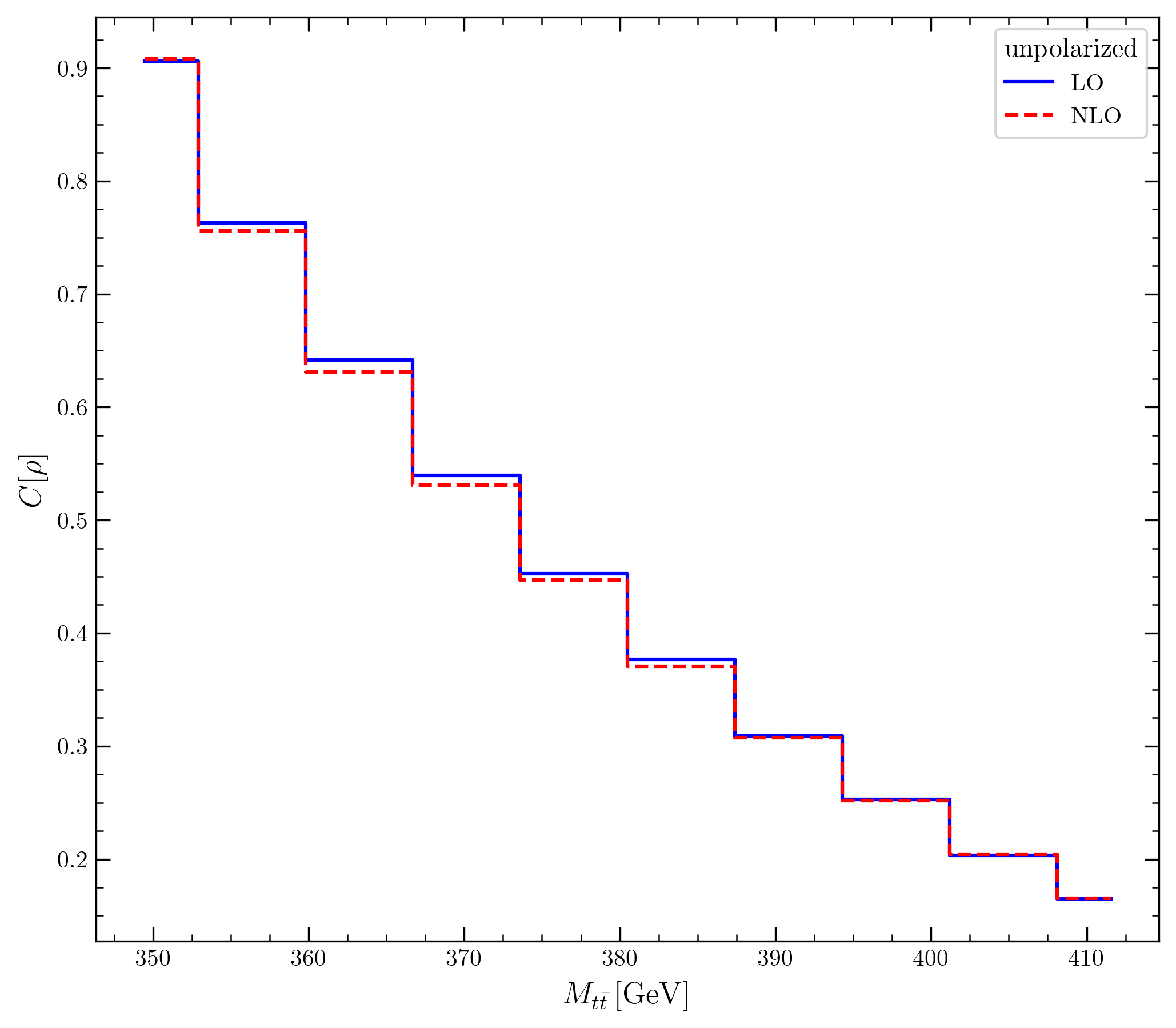} }
		\hspace{-0.3cm}~
		\subfigure[]{\label{m12_unpol}
			\includegraphics[width=0.4\textwidth]{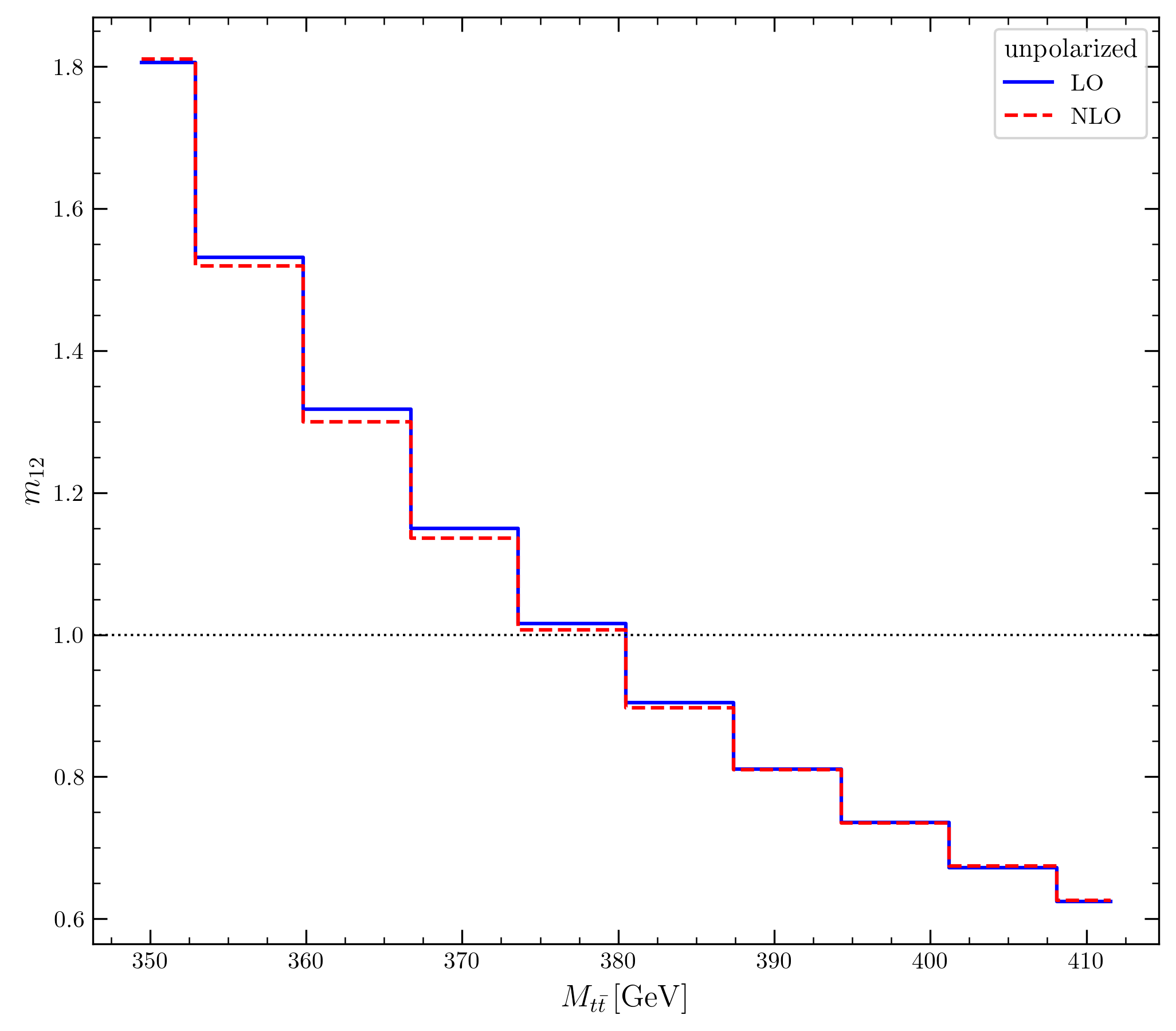} }	
\caption{
Invariant-mass distributions of the concurrence \(C[\rho]\) and the Bell nonlocality parameter \(m_{12}\) for the unpolarized configuration at \(\sqrt{s}=500~\mathrm{GeV}\). 
}
	\label{fig:unpol}
	\end{center}
\end{figure}

For the unpolarized initial state at \(\sqrt{s}=500~\mathrm{GeV}\), shown in Fig.~\ref{fig:unpol}, both the concurrence \(C[\rho]\) and the Bell nonlocality parameter  \(m_{12}\)  decrease with increasing invariant mass \(M_{t\bar t}\).
A direct comparison with Ref.~\cite{Choi:2026omc} is not straightforward, since their results depend on both invariant masses and the scattering angle, whereas ours are inclusive over the scattering angle.
The concurrence reaches its maximum value near the production threshold and decreases gradually with increasing $M_{t\bar t}$. 
It remains nonzero over the entire kinematic range, indicating that the $t\bar t$ system remains  entangled throughout the accessible phase space.
The Bell nonlocality parameter $m_{12}>1$ in the low-$M_{t\bar t}$ region, indicating the presence of Bell-inequality violation. With increasing $M_{t\bar t}$, $m_{12}$ decreases steadily and eventually becomes smaller than unity, implying that Bell-inequality violation is no longer observed in the high-$M_{t\bar t}$ region.
The NLO QCD corrections are generally small and lead only to minor quantitative modifications, without changing the  behavior of either \(C[\rho]\) or \(m_{12}\) as functions of \(M_{t\bar t}\).
We have also checked the sensitivity to real-emission phase-space restrictions by imposing transverse-momentum cuts on the emitted gluon. 
The impact is most visible near threshold, while the overall behavior remains unchanged.

\begin{figure}[htbp]
	\begin{center}
			\includegraphics[width=0.9\textwidth]{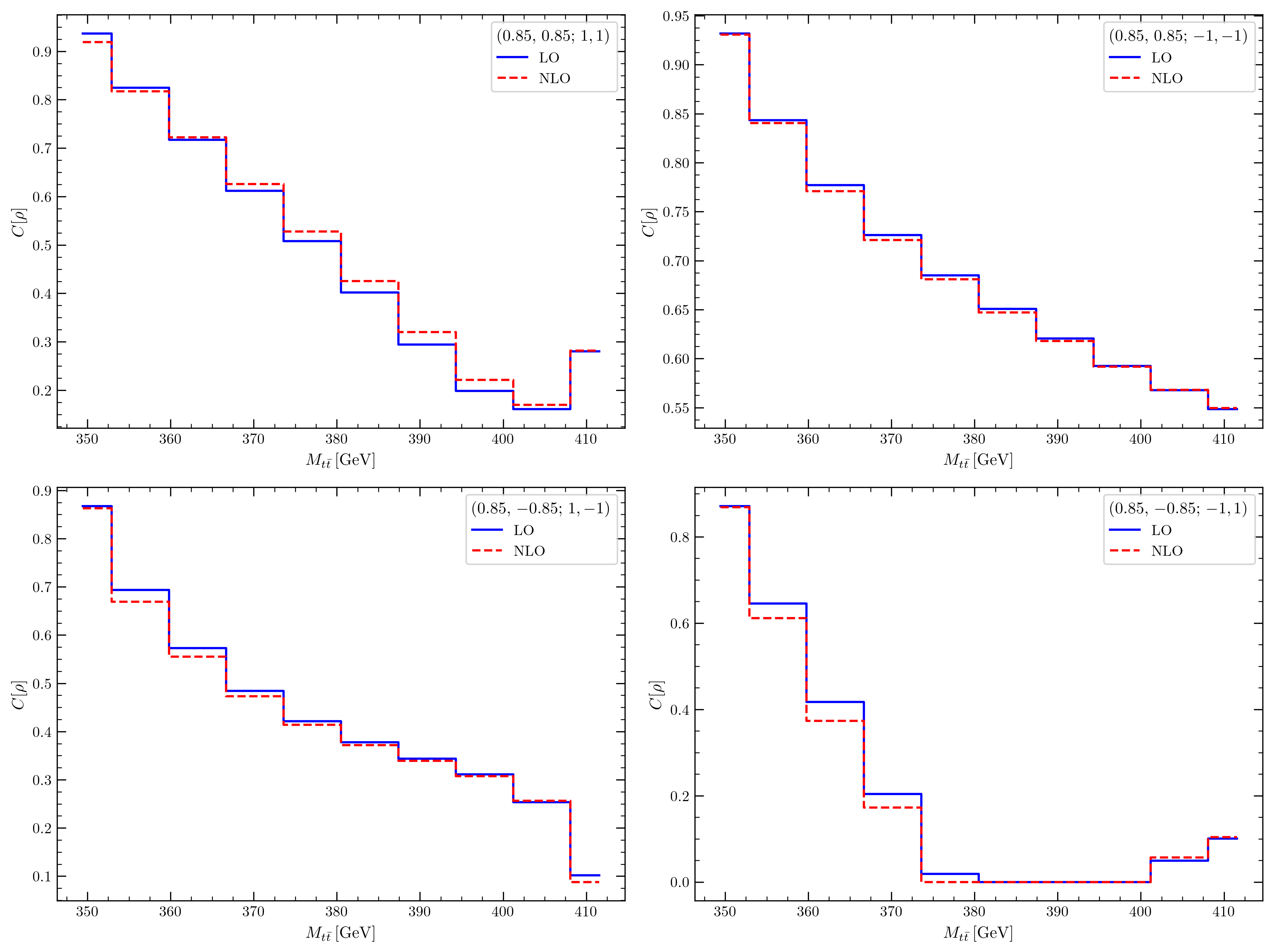} 	
\caption{Invariant-mass distributions of the concurrence  \(C[\rho]\)  under polarized  configuration at \(\sqrt{s}=500~\mathrm{GeV}\). 
The upper-left panel corresponds to the configuration
$(P_e^{(1)}, P_e^{(2)} ; P_L^{(1)}, P_L^{(2)})=(0.85,0.85;1,1)$,
the upper-right panel to
$(0.85,0.85;-1,-1)$,
the lower-left panel to
$(0.85,-0.85;1,-1)$,
and the lower-right panel to
$(0.85,-0.85;-1,1)$.
}
	\label{fig:concurrence_4panel}
	\end{center}
\end{figure}
For  polarized configurations at \(\sqrt{s}=500~\mathrm{GeV}\), the concurrence $C[\rho]$ as shown in Fig.~\ref{fig:concurrence_4panel} presents a strong dependence on the initial-state polarization, indicating that beam polarization has a significant impact on the quantum entanglement of the $t\bar t$ system. 
At LO, the overall trends for polarized initial states are qualitatively similar to those reported in Ref.~\cite{Choi:2026omc}, despite the different polarization configurations and scattering angle treatments.
All configurations exhibit sizable entanglement near the production threshold, while their evolution with increasing invariant mass $M_{t\bar t}$ differs substantially. 
For the configuration $(P_e^{(1)}, P_e^{(2)} ; P_L^{(1)}, P_L^{(2)})=(0.85,0.85;1,1)$, $C[\rho]$ is relatively large in the low-invariant-mass region and gradually decreases with increasing $M_{t\bar t}$, with a slight enhancement reappearing above  $M_{t\bar t} \simeq 410$  GeV. 
However, the $(0.85,0.85;-1,-1)$ configuration exhibits the strongest entanglement enhancement, where $C[\rho]$ remains sizable throughout the entire kinematic range and decreases much more slowly, indicating that this polarization setup can sustain strong quantum entanglement over a broad invariant-mass region. 
In the $(0.85,-0.85;1,-1)$ configuration, $C[\rho]$ decreases continuously as $M_{t\bar t}$ increases. 
For the $(0.85,-0.85;-1,1)$ configuration, $C[\rho]$ becomes  zero in the intermediate invariant-mass region and reappears  in the high-invariant-mass region, exhibiting a more pronounced nonmonotonic behavior. 
The nonmonotonic behavior observed in this polarized configuration  can be understood from the change in the spin correlation structure across different invariant-mass regions. 
Near the production threshold, the spin correlation is dominated by the \(k\)-component, which drives the $t\bar t$ state closer to a Bell-like state and therefore enhances the concurrence. 
As \(M_{t\bar t}\) increases, the spin correlation structure gradually changes, leading to a significant suppression of entanglement in the intermediate-\(M_{t\bar t}\) region, while in the large-\(M_{t\bar t}\) region, the concurrence becomes nonzero again, resulting in the observed nonmonotonic behavior.
The NLO QCD corrections to  $C[\rho]$  are generally small for all polarized configurations, leading only to minor quantitative modifications without altering the  behavior of these observables.

\begin{figure}[htbp]
	\begin{center}
			\includegraphics[width=0.9\textwidth]{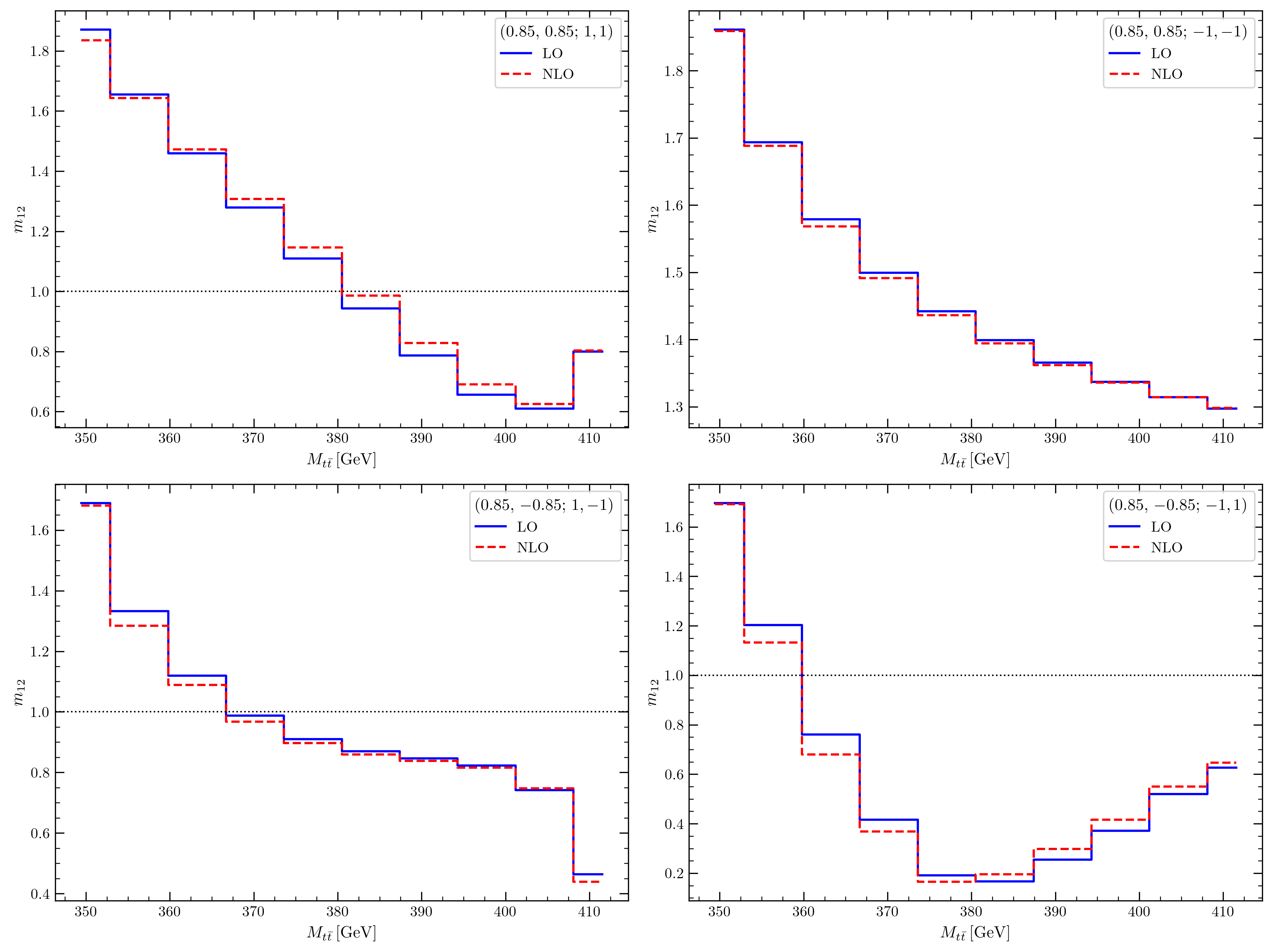} 	
\caption{Invariant-mass distributions of the Bell nonlocality parameter \(m_{12}\) under polarized  configuration at \(\sqrt{s}=500~\mathrm{GeV}\). 
The upper-left panel corresponds to the configuration
$(P_e^{(1)}, P_e^{(2)} ; P_L^{(1)}, P_L^{(2)})=(0.85,0.85;1,1)$,
the upper-right panel to
$(0.85,0.85;-1,-1)$,
the lower-left panel to
$(0.85,-0.85;1,-1)$,
and the lower-right panel to
$(0.85,-0.85;-1,1)$.
}
	\label{fig:m12_4panel}
	\end{center}
\end{figure}
The Bell nonlocality parameter $m_{12}$ as shown in Fig.~\ref{fig:m12_4panel} exhibits even more pronounced differences among the polarized configurations. 
For the $(0.85,0.85;1,1)$ configuration, $m_{12}$ is significantly larger than unity in the low-invariant-mass region, but gradually decreases below unity as $M_{t\bar t}$ increases, with only a slight enhancement above  $M_{t\bar t} \simeq 410$  GeV. 
For the $(0.85,0.85;-1,-1)$ configuration, the Bell-type correlation is the strongest, with $m_{12}>1$ throughout the entire considered $M_{t\bar t}$ range, indicating persistent Bell-inequality violation over the full kinematic region. 
In the $(0.85,-0.85;1,-1)$ configuration, $m_{12}>1$ occurs only in a limited region near threshold before  dropping below unity. 
In the $(0.85,-0.85;-1,1)$ configuration, although Bell-inequality violation is present near threshold, it is rapidly suppressed at higher energies; even when the concurrence becomes nonzero again in the high-invariant-mass region, $m_{12}$ remains below unity. 
The NLO QCD corrections to   $m_{12}$ do not change its characteristic dependence on $M_{t\bar t}$, preserving the qualitative pattern of Bell correlations for all considered polarization setups.

It is useful to further compare the NLO QCD effects on the total production rate and on the quantum observables. 
As shown above, the NLO corrections can substantially enhance the total $t\bar t$ production cross section, while their effects on the spin correlations and quantum observables are relatively moderate. 
This difference is a consequence of the normalized nature of the spin observables. 
The quantum observables considered in this work are constructed from the normalized spin density matrix, with 
\[
	\rho = \frac{R}{\mathrm{Tr}[R]},
\qquad
B_i^\pm = \frac{\tilde B_i^\pm}{\tilde A}, 
\qquad
C_{ij} = \frac{\tilde C_{ij}}{\tilde A}.
\]
The NLO QCD corrections to the numerators and denominators are largely suppressed in the ratios.
As a result, the concurrence $C[\rho]$ and the Bell nonlocality parameter $m_{12}$ exhibit substantially weaker sensitivity radiative corrections than the total cross section in the present inclusive parton-level calculation.

\section{Summary}\label{sec5}

In this work, we have investigated the spin correlations and quantum entanglement  of the process $\gamma\gamma\to t\bar t$ at a Compton backscattering photon collider and presented a systematic study of the corresponding NLO QCD corrections for various beam-polarization configurations.

The production rate of $t\bar t$ pairs exhibits a strong dependence on the initial beam polarizations.
The NLO QCD corrections provide a positive and relatively stable enhancement to the total cross sections, corresponding to $K$-factors in the range $1.4$--$1.6$.
For spin observables, the single-particle polarization vanishes  at LO for unpolarized initial states, while polarized beams generate polarization effects of top quark. 
Spin correlation $C_{kk}$ exhibits the maximum correlation in most polarization configurations. 
We find that NLO QCD corrections to the spin observables remain at the few-mille level.

We also find nontrivial behaviors in the quantum observables. The concurrence is strongest near the production threshold and gradually decreases with increasing $M_{t\bar t}$.
In particular, it remains sizable over a broad $M_{t\bar t}$ range for the $(0.85,0.85;-1,-1)$ configuration, whereas it vanishes in the intermediate invariant-mass region for the $(0.85,-0.85;-1,1)$ configuration.
In addition, the Bell nonlocality parameter $m_{12}>1$ persists over the entire investigated range for the $(0.85,0.85;-1,-1)$ configuration, whereas it exceeds unity only in the low-$M_{t\bar{t}}$ region for most polarization configurations.
The results presented in this work provide a solid theoretical basis for future studies of quantum correlations in top-quark pair production at photon colliders.

\section*{Acknowledgements}

The authors thank  the members of the Institute of Theoretical Physics of Shandong University for their helpful communications. 
This work is supported in part by National Natural Science Foundation of China under the Grants  No.12321005,  No.12235008 and No.12475083.
\newpage

\bibliography{ref}

\end{document}